\documentclass[12pt]{jpconf}
\pdfoutput=1
\usepackage{graphicx}
\usepackage{footmisc}
\usepackage{color}
\usepackage{amsmath} 
\usepackage{array,relsize,float}
\usepackage[bitstream-charter]{mathdesign}

\renewcommand{\thefigure}{\arabic{figure}}

\def\nn{\nonumber}

\def\beq{\begin{equation}}
\def\eeq{\end{equation}}
\def\bea{\begin{eqnarray}}
\def\eea{\end{eqnarray}}
\def\beqn{\begin{eqnarray}} 
\def\eeqn{\end{eqnarray}}
\def\nn{\nonumber}

\def\qon#1{q_{#1,0}^{(+)}}

\def\ket#1{|{#1}\rangle}

\renewcommand{\thefootnote}{\fnsymbol{footnote}}

\begin{document}
\title{Geometrical causality: casting Feynman integrals into quantum algorithms}

\author{G. F. R. Sborlini$^{1,2}$\footnote[1]{\hspace{1mm}Speaker}}

\address{$^1$ Departamento de F\'isica Fundamental e IUFFyM, Universidad de Salamanca, 
37008 Salamanca, Spain.}
\address{$^2$ Escuela de Ciencias, Ingenier\'ia y Diseño, Universidad Europea de Valencia, \\ Paseo de la Alameda 7, 46010 Valencia, Spain.}

\ead{german.sborlini@usal.es}

\begin{abstract}
The calculation of higher-order corrections in Quantum Field Theories is a challenging task. In particular, dealing with multiloop and multileg Feynman amplitudes leads to severe bottlenecks and a very fast scaling of the computational resources required to perform the calculation. With the purpose of overcoming these limitations, we discuss efficient strategies based on the Loop-Tree Duality, its manifestly causal representation and the underlying geometrical interpretation. In concrete, we exploit the geometrical causal selection rules to define a Hamiltonian whose ground-state is directly related to the terms contributing to the causal representation. In this way, the problem can be translated into a minimization one and implemented in a quantum computer to search for a potential speed-up.
\end{abstract}

\setcounter{page}{1}

\section{Introduction}
\label{sec:Introduction}
It is a well-known fact that particle physics is going through the \emph{precision era}. Most of the theoretical predictions extracted from the Standard Model (SM) seem compatible with the experimental data, within the estimated errors. Then, reducing the errors (both from experiment and theory) will allow to test the fundamental parameters of SM and shed light into potential new physics signals. 
From the theory side, reaching more accurate predictions forces us to deal with complex calculations. In particular, in the context of high-energy physics (HEP), this implies the need of higher-orders in perturbation theory. Then, we must compute multiloop Feynman amplitudes and integrate multiparticle phase-spaces: both operations are plagued of technical bottlenecks that prevent a straightforward calculation.

In the last decades, a tremendous progress has been done regarding Feynman integral calculations\footnote{We refer the interested reader to the very complete review presented in Ref. \cite{Heinrich:2020ybq}.}. Even if several drawbacks were solved, breaking the current precision frontier leads to expressions that are so complicated that they cannot be efficiently handled by traditional methods. 
In the direction of exploring new strategies for tackling Feynman integrals and scattering cross-sections, the Loop-Tree Duality (LTD) \cite{Catani:2008xa,Bierenbaum:2010cy,Bierenbaum:2012th,Buchta:2014dfa,Tomboulis:2017rvd,Buchta:2015xda} was unveiled to connect loop and phase-space calculations in a natural way. The underlying idea of the formalism is to remove the energy component of loop momenta, so that the remaining integration is performed in a Euclidean space, resembling a traditional phase-space integration\footnote{A more complete review about the LTD and its development is available in Ref. \cite{deJesusAguilera-Verdugo:2021mvg}.}. With the time, this framework has evolved to allow efficient asymptotic expansions \cite{Driencourt-Mangin:2017gop,Plenter:2019jyj,Plenter:2020lop,Plenter:2022zxk}, local numerical renormalization \cite{Driencourt-Mangin:2019aix,Driencourt-Mangin:2019yhu} and local integrand-level representations of benchmark NLO cross-sections \cite{Hernandez-Pinto:2015ysa,Sborlini:2016gbr,Sborlini:2016hat,Prisco:2020kyb}, among other applications in HEP.

Furthermore, the LTD framework was recently reformulated by exploiting the emergence of a causal representation from the iterated application of Cauchy's residue theorem \cite{Aguilera-Verdugo:2019kbz,Verdugo:2020kzh,Aguilera-Verdugo:2020kzc,Aguilera-Verdugo:2020nrp,Ramirez-Uribe:2020hes,Ramirez-Uribe:2022sja,Runkel:2019yrs,Runkel:2019zbm,Capatti:2019edf,Capatti:2019ypt}. This representation can be derived from algebraic \cite{TorresBobadilla:2021ivx,TorresBobadilla:2021dkq} and geometrical \cite{Sborlini:2021owe,Capatti:2022mly} approaches, thus avoiding an explicit handling of the resulting expressions from the nested residue calculation. In particular, we have recently shown that the geometrical formalism is suitable to study the causal representation of multiloop scattering amplitudes using quantum computers \cite{Ramirez-Uribe:2021ubp,Clemente:2022nll}, which might allow a faster calculation. 

In this article, we explain how the geometrical causal representation of LTD amplitudes is specially suited to implement Feynman integral calculations in quantum devices. In Sec. \ref{sec:CausalLTD} we present a brief explanation of the LTD framework, with special emphasis in the causal representation. Then, in Sec. \ref{ssec:Geometry}, the connection between graph theory and causality is outlined through the introduction of the geometrical causal selection rules. Making use of this geometrical concepts, we explain in Sec. \ref{sec:QAcausal} how the causal representation can be bootstrapped from the collection of directed acyclic graphs (DAGs), and how to detect these graphs with quantum algorithms. In particular, in Sec. \ref{ssec:VQEHamiltonian}, we explain how to build a Hamiltonian whose ground-state contains all the possible DAGs, and how to find the configurations with minimum energy using VQE-based algorithms. Finally, in Sec. \ref{sec:conclusions}, we present the conclusions and discuss possible future research directions.

\section{Causal Loop-Tree Duality}
\label{sec:CausalLTD}
As mentioned in the introduction, the LTD formalism allows to decrease one degree of freedom per loop integration when considering multiloop multileg scattering amplitudes. If the energy component is removed, then the resulting integral is defined over an Euclidean space (instead of a Minkowski one), which transforms loops into phase-space integrals. So, let us start from a generic $L$-loop scattering amplitude with $P$ external particles in the Feynman representation,
\beq
{\cal A}_F^{(L)} = \int_{\ell_1 \ldots \ell_L} {\cal N} \big(\{\ell_s\}_L, \{p_j\}_P\big) \prod_{i=1}^n G_F(q_i)~,
\label{eq:MultiloopFeynman}
\eeq
where $q_i$ with $i\in \{1,\ldots,n\}$ are the momenta flowing through each Feynman propagator, $G_F(q_i)=(q_i^2-m_i^2+ \imath 0)^{-1}$ is the Feynman propagator and ${\cal N}$ represents a generic numerator. By re-writing the Feynman propagators as
\beq
G_F(q_i)= \frac{1}{q_{i,0}-\qon{i}} \times \frac{1}{q_{i,0}+\qon{i}} \, ,
\label{eq:GFconplus}
\eeq
with $\qon{i} = \sqrt{\vec{q_i}^2+m_i^2- \imath 0}$ the positive on-shell energy associated to the $i$-th internal line, we can compute the nested residues \cite{Verdugo:2020kzh} and add all the terms to obtain \cite{Aguilera-Verdugo:2020kzc,Aguilera-Verdugo:2020nrp}
\beqn
{\cal A}_D^{(L)} &=& \int_{\vec \ell_1 \ldots \vec \ell_L} 
\frac{1}{x_n} \sum_{\sigma  \in \Sigma} {\cal N}_{\sigma}\, \prod_{i=1}^{n-L} \, \frac{1}{\lambda_{\sigma(i)}^{h_{\sigma(i)}}} \
+ (\lambda^+_p \leftrightarrow \lambda^-_p)~,
\label{eq:CausalMaster}
\eeqn
with $h_{\sigma(i)} = \pm 1$. In the previous expression, $\lambda_{\sigma(i)}^{h_{\sigma(i)}}$ are called \emph{causal thresholds} and they encode any possible \emph{physical} threshold singularity of the underlying loop scattering amplitude: $1/\lambda$ are the so-called \emph{causal propagators}. It is worth noticing that the causal thresholds only involve sums of positive on-shell energies, i.e.
\beq
\lambda_{\sigma(i)}^{h_{\sigma(i)}} = \sum_{p \in O_i} \qon{p} \pm P_{i,0} \, ,
\eeq
with $P_{i,0}$ a combination of external momenta energies. The quantity $n$ in Eq. (\ref{eq:CausalMaster}) represents the number of internal momenta sets: two internal momenta $q_i$ and $q_j$ are said to belong to the set $s$ if they depend on the same linear combination of primitive loop momenta $\{\ell_r\}$. $k=n-L$ is known as the \emph{order} of the diagram, and it tells us how many causal thresholds \emph{must} be simultaneously entangled to reproduce all the possible threshold singularities of the diagram \cite{Sborlini:2021owe,Capatti:2022mly}: then, $\Sigma$ indicates the set of all the allowed causal entangled thresholds. In this way, Eq. (\ref{eq:CausalMaster}) is a generalization of the well-known Cutkosky rules \cite{Cutkosky:1960sp} and enables the reconstruction of the complete amplitude. 

Finally, let us notice that the factor $x_n = \prod_n 2\qon{i}$ transforms the integration measure from $d^{d}\ell_i$ into $\approx d^{d-1}\vec{\ell}_i / (2 E_i)$, namely the phase-space measure with the proper normalization. For this reason, the causal LTD formalism is specially suited to tackle the cross-section calculation at higher-orders in a completely unified way, without splitting into real (extended phase-space integration) and virtual (extended loop integration) corrections \cite{Rodrigo:2023PREP}.

\subsection{Geometry, causality and causal propagators}
\label{ssec:Geometry}
In order to achieve a purely geometrical description of the causal LTD representation, it is necessary to introduce some previous concepts inspired in graph theory. The first observation is that Feynman diagrams are \emph{oriented graphs} made of \emph{vertices} (codifying the interaction among particles) and \emph{edges} (associated to the propagation of virtual states). We define \emph{reduced Feynman graphs} by collapsing all the edges connecting a certain pair of vertices into a single \emph{multi-edge}. For the sake of simplicity, in this article, we work at the level of reduced graphs, so we can unambiguously refer to \emph{multi-edges} as simply \emph{edges}. The way in which vertices $V$ are connected through a set of edges $E$ is encoded within the adjacency matrix $A$: $(A)_{ij}$ is 1 ($-1$) if an edge from $i$ to $j$ ($j$ to $i$) exists, otherwise it is 0. It is important to highlight that the causal structure of a given Feynman diagram is completely determined by its reduced graph and, hence, by its adjacency matrix \cite{Sborlini:2021owe,Sborlini:2021nqu}. Moreover, we can proof that the order of a diagram, $k=n-L$, directly related to the number of vertices, namely $k=V-1$ \cite{Sborlini:2021owe,Capatti:2022mly}.

Given a reduced Feynman graph, we can built all the binary partitions of connected vertices, ${\cal P}_V^C$. These binary partitions codify all allowed physical thresholds, which originate when the diagram is split into two consistent connected subgraphs. This condition is directly motivated by Cutkosky's rules \cite{Cutkosky:1960sp}. In this way, given a partition $p$, we define the \emph{conjugated causal thresholds}, $\bar{\lambda}_{p}$, by summing all the momenta flowing through the partition. These conjugated causal thresholds are in one-to-one correspondence with all the possible $\lambda$'s appearing in the denominator of Eq. (\ref{eq:CausalMaster}).

So, by using only geometrical concepts, we manage to identify all the elements appearing in the causal LTD master formula. The remaining ingredient is the recipe for properly entangling the causal thresholds. Summarizing Ref. \cite{Sborlini:2021owe}, there are three compatibility rules to be checked:
\begin{enumerate}
 \item All the edges involved in a causal threshold must carry momenta flowing in the same direction.
 \item Given a combination of causal entangled thresholds, all the edges must be cut at least once. This means that the product $\prod_{i=1}^{k} \, \lambda_{\sigma(i)}^{h_{\sigma(i)}}$ for every allowed $\sigma \in \Sigma$ depends on $\qon{j}$ for all $j \in E$.
 \item Causal thresholds do no intersect: if we represent each $\lambda_p$ by a line delimiting the partition $p$, then these lines do not cross each other.
\end{enumerate}
The first condition turns out to be equivalent to order the momenta of all the edges in such a way that there are no cycles (loops). In other words, it implies that the graphical representation of a causal entangled threshold corresponds to a DAG \cite{Ramirez-Uribe:2021ubp}. 

The geometrical compatibility rules also suggest a procedure to build the causal representation. Given a reduced Feynman graph, one possibility consists in identifying all the associated DAGs. Then, we generate all the causal thresholds $\{\lambda_p\}$ by computing all the possible binary connected partitions. After that, we dress the previously identified DAGs with $k=V-1$ different causal propagators in such a way that the resulting combination fulfills (ii) and (iii). Within this strategy, the efficient identification of all the possible DAGs of a given Feynman diagram is a crucial step for bootstrapping a causal LTD representation. 

\section{Quantum algorithms and causal flow}
\label{sec:QAcausal}
When dealing with multileg multiloop scattering amplitudes, the number of vertices and edges of the underlying Feynman diagrams scales very fast. As a consequence, the complexity of the causal representation grows almost \emph{exponentially} because of the large number of allowed entangled causal thresholds. In particular, having in mind the recipe for geometrical reconstruction, at a certain point we would need to identify all the possible DAGs of a given reduced Feynman diagram. Since the number of directed graphs is $2^E$ with $V \leq E \leq V(V-1)/2$, there is an exponential number of configurations to be tested for the acyclicity condition. For solving this task, there are constructive \emph{classical} algorithms which perform well for sparse graphs. However, when the graphs are dense or maximally-connected, the complexity is exponential.

With this panorama in mind, and noticing that the identification of DAGs is the first step towards the full causal reconstruction, we decided to explore alternative algorithms in search of a potential speed-up. One of these alternatives comprises quantum algorithms. They have been very recently applied to a wide variety of problems in HEP, including jet reconstruction~\cite{Wei:2019rqy,Pires:2021fka,Pires:2020urc,deLejarza:2022bwc,Delgado:2022snu,Barata:2022wim,Barata:2021yri}, determination of parton densities (PDFs)~\cite{Perez-Salinas:2020nem}, anomaly detection~\cite{Ngairangbam:2021yma}, integration of scattering processes~\cite{Agliardi:2022ghn,deLejarza:2023qxk}, among others. In our case, we decided to implement a Grover-based search algorithm \cite{Grover:1997ch,Grover:1997fa} to identify all the DAGs within a given Feynman diagram. In Ref. \cite{Ramirez-Uribe:2021ubp}, we codify the direction of each edge into a single qubit. This choice is motivated by Eq. (\ref{eq:GFconplus}), since a propagator in the Feynman representation can be understood as a superposition of a particle traveling forward and backward in time.

\begin{figure}[h]
\begin{center}
   \includegraphics[width=0.8\linewidth]{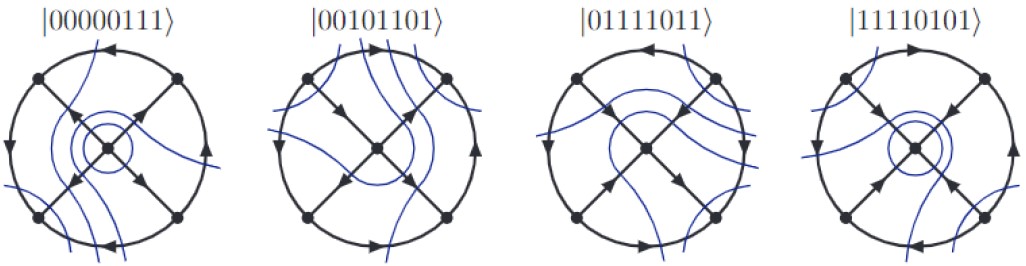}
    \includegraphics[width=0.9\linewidth]{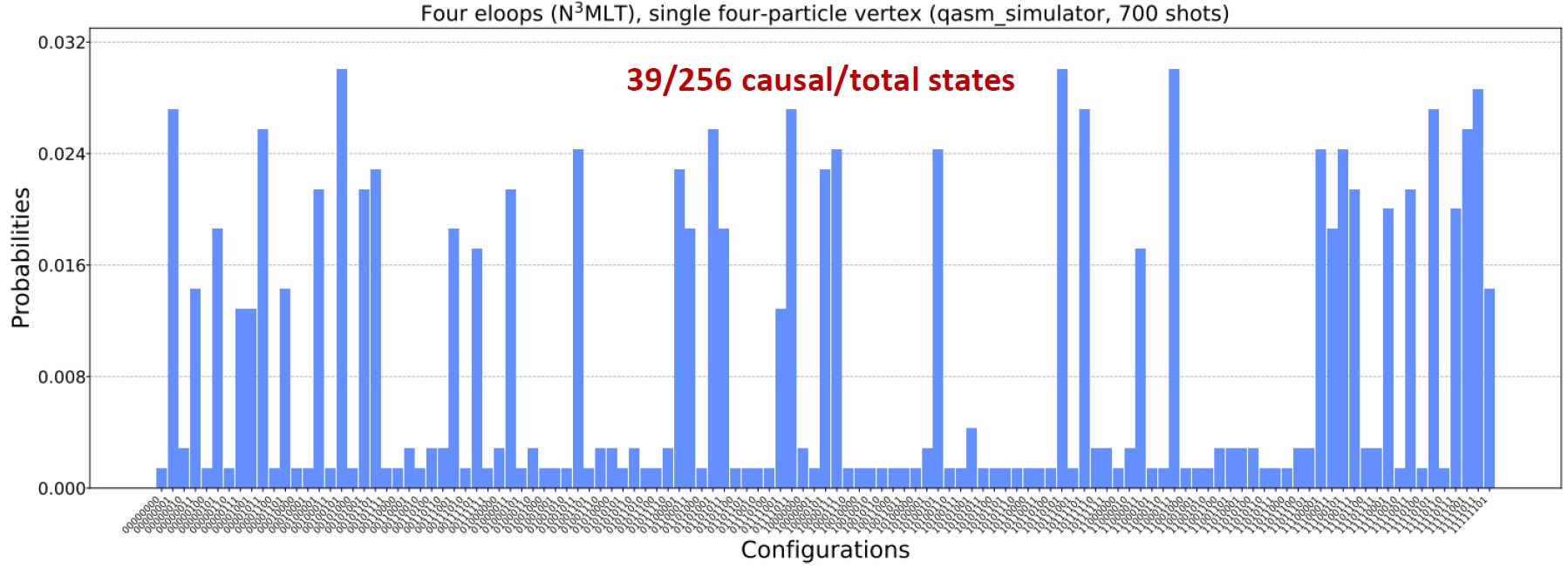}
\end{center}
\caption{Representative causal entangled thresholds for a four-loop N$^3$MLT or pizza topology (upper plot). Probability distribution originated by our Grover-based DAG search algorithm that allow to identify the 39 acyclic configurations (out of 256 possible states) with a 100 $\%$ success rate (lower plot).}
\label{fig:Figura1}
\end{figure}

In Fig. \ref{fig:Figura1}, we present the result of the application of our algorithm to a four-loop N$^3$MLT or pizza topology, which contains eight edges and five vertices. In the upper part of the figure, we present four representative cyclic configurations dressed with the appropriate causal entangled thresholds (following the recipe presented in Sec. \ref{ssec:Geometry}). The corresponding quantum circuit was implemented in a \texttt{Qiskit} simulator, and required 35 qubits (including ancillary ones) with transpiled depth 55. After 700 shots, we reach a 5-sigma discrimination among cyclic and acyclic states, which allows to reach a 100 $\%$ success rate in the identification of DAGs.

\subsection{Hamiltonian formulation for VQE}
\label{ssec:VQEHamiltonian}
The Grover-based algorithm turned out to be very efficient and provide a certain speed-up w.r.t. classical strategies. Although the success rate was perfect with a very reduced number of shots, the rapid growth in the number of qubits and the need of implementing depth circuits translates into a large consumption of quantum resources, which are not available in current devices. Thus, we decided to explore another strategy to identify DAGs: the minimization of a Hamiltonian using Variational Quantum Eigensolvers (VQE).

\begin{figure}[h]
\begin{center}
   \includegraphics[width=0.4\linewidth]{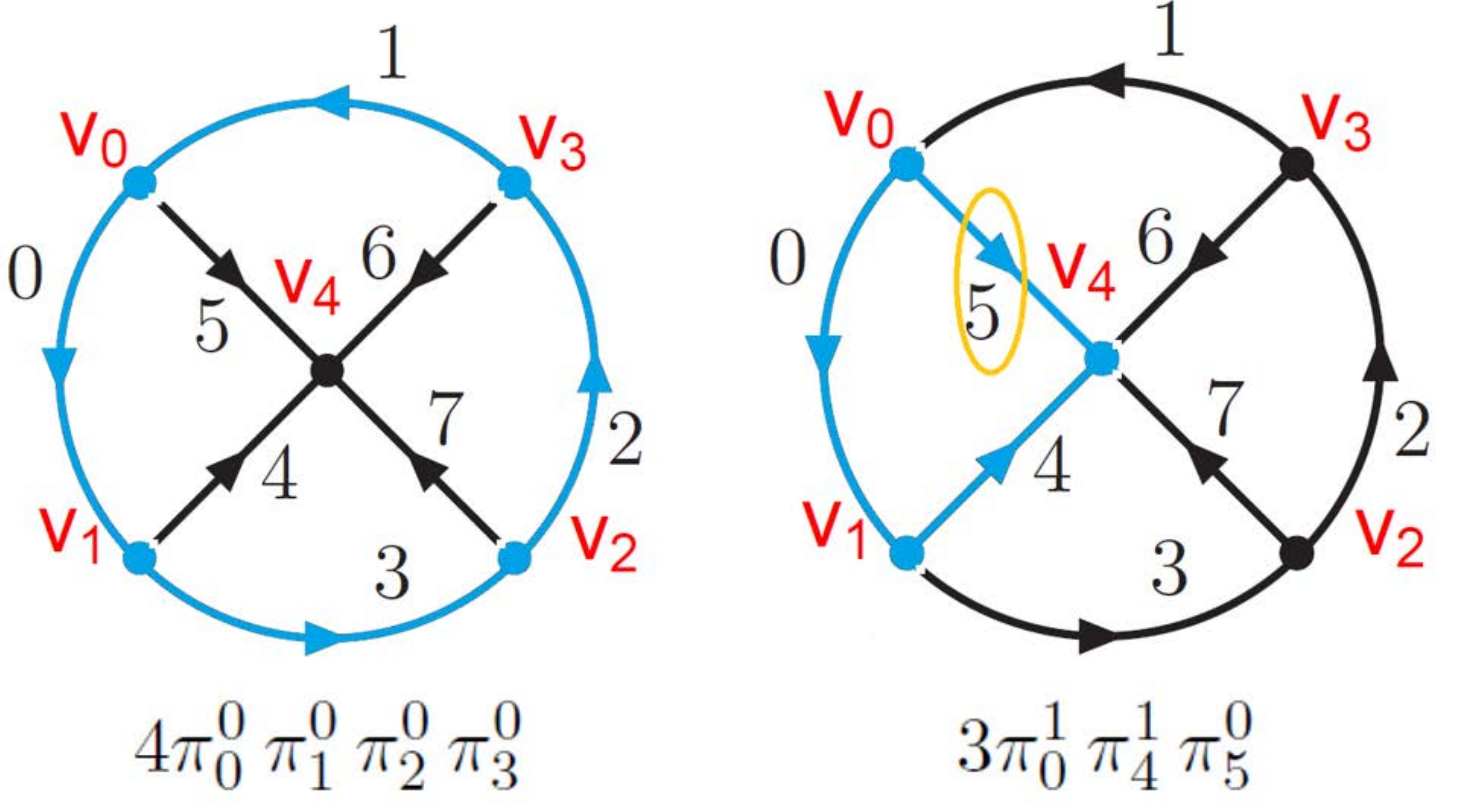}
\end{center}
\caption{Representation of two terms contributing to the loop Hamiltonian of a pizza topology. In blue, we indicate the edges involved in the loop, being the direction of each arrow the associated to $G_0$. In the triangle (left figure), we see that $e_5$ must be reversed in order to have a closed loop, which is reflected by the inverted projector $\pi_5^0$.}
\label{fig:Figura2}
\end{figure}

The first part of this strategy consisted in building a Hamiltonian whose ground-states embodies all the DAGs of a given graph. In order to do so, we rely on the adjacency matrix. Explicitly, we start by fixing an arbitrary orientation of all the edges of the graph; we denote by $G_0=(E_0,V)$ this oriented graph. We define $e_i\approx\ket{0}$ ($e_i\approx\ket{1}$) if the direction of the edge $e_i$ is the same (opposite) as in $G_0$. Then, we introduce the projector operator in a single edge, $\pi_i^{(0,1)}$, defined as $\pi_i^0 \ket{\psi} = \ket{\psi}$ and $\pi_i^1 \ket{\psi} = 0$ if $e_i$ is in the same direction as in $G_0$. With all of these, it is possible to write the adjacency matrix $A$ as a function of $\pi$ operators and we define the Hamiltonian as
\beq
H_{G_0} = \sum\limits_{n=1}^{M_{G_0}} \text{tr} (A^n)  \, ,
\label{eq:Hamiltonian}
\eeq
with $M_{G_0}$ the maximal size of the cycles in $G_0$. The ground-state is composed by all DAGs and has energy 0. For instance, if we consider a pizza or N$^3$MLT topology, the Hamiltonian reads
\beqn
\nn H_{\rm N^3MLT} &=& 4\pi_{0}^0\,  \pi_{1}^0\,  \pi_{2}^0\,  \pi_{3}^0\,+4\pi_{0}^1\,  \pi_{1}^1\,  \pi_{2}^1\,  \pi_{3}^1\,+3\pi_{0}^1\,  \pi_{4}^1\,  \pi_{5}^0\,+5\pi_{1}^0\,  \pi_{2}^0\,  \pi_{3}^0\,  \pi_{4}^1\,  \pi_{5}^0\,+3\pi_{0}^0\,  \pi_{4}^0\,  \pi_{5}^1\,
\\ \nn &+& 5\pi_{1}^1\,  \pi_{2}^1\,  \pi_{3}^1\,  \pi_{4}^0\,  \pi_{5}^1\,+4\pi_{0}^1\,  \pi_{1}^1\,  \pi_{4}^1\,  \pi_{6}^0\,+4\pi_{2}^0\,  \pi_{3}^0\,  \pi_{4}^1\,  \pi_{6}^0\,+3\pi_{1}^1\,  \pi_{5}^1\,  \pi_{6}^0\,+5\pi_{0}^0\,  \pi_{2}^0\,  \pi_{3}^0\,  \pi_{5}^1\,  \pi_{6}^0\,
\\ \nn &+& 4\pi_{0}^0\,  \pi_{1}^0\,  \pi_{4}^0\,  \pi_{6}^1\,+4\pi_{2}^1\,  \pi_{3}^1\,  \pi_{4}^0\,  \pi_{6}^1\,+3\pi_{1}^0\,  \pi_{5}^0\,  \pi_{6}^1\,+5\pi_{0}^1\,  \pi_{2}^1\,  \pi_{3}^1\,  \pi_{5}^0\,  \pi_{6}^1\,+5\pi_{0}^1\,  \pi_{1}^1\,  \pi_{2}^1\,  \pi_{4}^1\,  \pi_{7}^0\,
\\ \nn&+& 3\pi_{3}^0\,  \pi_{4}^1\,  \pi_{7}^0\,+4\pi_{1}^1\,  \pi_{2}^1\,  \pi_{5}^1\,  \pi_{7}^0\,+4\pi_{0}^0\,  \pi_{3}^0\,  \pi_{5}^1\,  \pi_{7}^0\,+3\pi_{2}^1\,  \pi_{6}^1\,  \pi_{7}^0\,+5\pi_{0}^0\,  \pi_{1}^0\,  \pi_{3}^0\,  \pi_{6}^1\,  \pi_{7}^0\,
\\ \nn&+& 5\pi_{0}^0\,  \pi_{1}^0\,  \pi_{2}^0\,  \pi_{4}^0\,  \pi_{7}^1\,+3\pi_{3}^1\,  \pi_{4}^0\,  \pi_{7}^1\,+4\pi_{1}^0\,  \pi_{2}^0\,  \pi_{5}^0\,  \pi_{7}^1\,+4\pi_{0}^1\,  \pi_{3}^1\,  \pi_{5}^0\,  \pi_{7}^1\,+3\pi_{2}^0\,  \pi_{6}^0\,  \pi_{7}^1\,
\\ &+& 5\pi_{0}^1\,  \pi_{1}^1\,  \pi_{3}^1\,  \pi_{6}^0\,  \pi_{7}^1 \, ,
\label{eq:HamiltonianoPizza}
\eeqn
which is defined w.r.t. the initial orientation draw in Fig. \ref{fig:Figura2}. In that figure, we present the graphical description of two concrete terms of $H$, to better explain their meaning.

Once the Hamiltonian is defined, we proceed to minimize it using a multi-run VQE strategy, which is an hybrid classical-quantum algorithm\footnote{For a complete review of VQE, we refer the interested reader to Ref. \cite{Tilly:2021jem}.}. In the original VQE, we start with an ansatz codified into a parameterized quantum circuit, and we measure the expectation value of the Hamiltonian on this ansatz. This feeds a classical optimizer which modifies the value of the parameters and the measure is repeated. The process is iterated till an approximation to the ground-state is found. This procedure works very well for problems with low-degeneration of the ground state. However, the identification of DAGs implies solving a multiple-degenerated problem (i.e. there could be up to $2^E$ global minima), so we executed multiple runs of the VQE, collect the solutions found and modify the Hamiltonian for the next run by adding penalization terms (to avoid re-finding the already detected solutions). By means of this strategy, the success rate jumped from ${\cal O}(4 \, \%)$ in the naive VQE to ${\cal O}(90 \, \%)$ with the improved multi-run VQE, as reported in Ref. \cite{Clemente:2022nll} with a set of benchmark topologies. 

\section{Conclusions and outlook}
\label{sec:conclusions}
In this article, we discussed how to use the Loop-Tree Duality (LTD) to cleverly rewrite Feynman integrals, switching from a Minkowski to a Euclidean integration domain. We proved that the nested residues strategy leads to a manifestly causal representation of multiloop scattering amplitudes. This causal representation can be re-derived by-passing the explicit calculation of the residues applying a set of geometrical rules. In concrete, given a reduced Feynman graph, we have to identify all the binary connected partitions (in one-to-one correspondence with the causal thresholds) and the associated directed acyclic graphs (DAGs). Then, the set of DAGs is \emph{dressed} with the causal propagators $\{\lambda_r\}$, in such a way that only a certain subset of \emph{compatible} causal entangled thresholds remain. The result of this procedure is Eq. (\ref{eq:CausalMaster}), namely a master formula describing the causal LTD representation of any multiloop multileg scattering amplitude.

Furthermore, it turned out that the geometrical formalism is specially suited for translating the reconstruction of causal representation into a problem of quantum computing. In fact, we have tackled the identification of DAGs by using Grover-based and multi-VQE algorithms, presenting proof-of-concepts with a high success rate. This is a very first and promising step towards a complete bootstrapping of causal representations with realistic quantum algorithms. Further developments in this direction (in particular, codifying the geometrical selection rules within a Hamiltonian) will have an important impact in the calculation of multiloop multileg scattering amplitudes, potentially allowing to surpass the current limitations of standard classical algorithms in quantum field theories.

\section{Acknowledgments}
I would like to thank G. Rodrigo, L. Vale Silva, A. Renter\'ia-Olivo (IFIC-Valencia), K. Jansen, A. Crippa, Y. Chai (DESY), R. Hern\'andez-Pinto and D. Renter\'ia-Estrada (UAS-Mexico) for fruitful comments and collaborating in the project. This work was partially supported by EU Horizon 2020 research and innovation program STRONG-2020 project under grant agreement No. 824093 and H2020-MSCA-COFUND-2020 USAL4EXCELLENCE-PROOPI-391 project under grant agreement No 101034371.

\section{References}

\bibliographystyle{JHEP}

\begin{thebibliography}{10}

\bibitem{Heinrich:2020ybq}
G.~Heinrich, \emph{{Collider Physics at the Precision Frontier}},
  \href{http://dx.doi.org/10.1016/j.physrep.2021.03.006}{\emph{Phys. Rept.}
  {\bf 922} (2021) 1--69}, [\href{http://arxiv.org/abs/2009.00516}{{\tt
  2009.00516}}].

\bibitem{Catani:2008xa}
S.~Catani, T.~Gleisberg, F.~Krauss, G.~Rodrigo and J.-C. Winter, \emph{{From
  loops to trees by-passing Feynman's theorem}},
  \href{http://dx.doi.org/10.1088/1126-6708/2008/09/065}{\emph{JHEP} {\bf 09}
  (2008) 065}, [\href{http://arxiv.org/abs/0804.3170}{{\tt 0804.3170}}].

\bibitem{Bierenbaum:2010cy}
I.~Bierenbaum, S.~Catani, P.~Draggiotis and G.~Rodrigo, \emph{{A Tree-Loop
  Duality Relation at Two Loops and Beyond}},
  \href{http://dx.doi.org/10.1007/JHEP10(2010)073}{\emph{JHEP} {\bf 10} (2010)
  073}, [\href{http://arxiv.org/abs/1007.0194}{{\tt 1007.0194}}].

\bibitem{Bierenbaum:2012th}
I.~Bierenbaum, S.~Buchta, P.~Draggiotis, I.~Malamos and G.~Rodrigo,
  \emph{{Tree-Loop Duality Relation beyond simple poles}},
  \href{http://dx.doi.org/10.1007/JHEP03(2013)025}{\emph{JHEP} {\bf 03} (2013)
  025}, [\href{http://arxiv.org/abs/1211.5048}{{\tt 1211.5048}}].

\bibitem{Buchta:2014dfa}
S.~Buchta, G.~Chachamis, P.~Draggiotis, I.~Malamos and G.~Rodrigo, \emph{{On
  the singular behaviour of scattering amplitudes in quantum field theory}},
  \href{http://dx.doi.org/10.1007/JHEP11(2014)014}{\emph{JHEP} {\bf 11} (2014)
  014}, [\href{http://arxiv.org/abs/1405.7850}{{\tt 1405.7850}}].

\bibitem{Tomboulis:2017rvd}
E.~Tomboulis, \emph{{Causality and Unitarity via the Tree-Loop Duality
  Relation}}, \href{http://dx.doi.org/10.1007/JHEP05(2017)148}{\emph{JHEP} {\bf
  05} (2017) 148}, [\href{http://arxiv.org/abs/1701.07052}{{\tt 1701.07052}}].

\bibitem{Buchta:2015xda}
S.~Buchta, \emph{{Theoretical foundations and applications of the Loop-Tree
  Duality in Quantum Field Theories}}.
\newblock PhD thesis, Valencia U., 2015.
\newblock \href{http://arxiv.org/abs/1509.07167}{{\tt 1509.07167}}.

\bibitem{deJesusAguilera-Verdugo:2021mvg}
J.~de~Jes\'us Aguilera-Verdugo et~al., \emph{{A Stroll through the Loop-Tree
  Duality}}, \href{http://dx.doi.org/10.3390/sym13061029}{\emph{Symmetry} {\bf
  13} (2021) 1029}, [\href{http://arxiv.org/abs/2104.14621}{{\tt 2104.14621}}].

\bibitem{Driencourt-Mangin:2017gop}
F.~Driencourt-Mangin, G.~Rodrigo and G.~F. Sborlini, \emph{{Universal dual
  amplitudes and asymptotic expansions for $gg\rightarrow H$ and $H\rightarrow
  \gamma \gamma $ in four dimensions}},
  \href{http://dx.doi.org/10.1140/epjc/s10052-018-5692-5}{\emph{Eur. Phys. J.
  C} {\bf 78} (2018) 231}, [\href{http://arxiv.org/abs/1702.07581}{{\tt
  1702.07581}}].

\bibitem{Plenter:2019jyj}
J.~Plenter, \emph{{Asymptotic Expansions Through the Loop-Tree Duality}},
  \href{http://dx.doi.org/10.5506/APhysPolB.50.1983}{\emph{Acta Phys. Polon. B}
  {\bf 50} (2019) 1983--1992}.

\bibitem{Plenter:2020lop}
J.~Plenter and G.~Rodrigo, \emph{{Asymptotic expansions through the loop-tree
  duality}},
  \href{http://dx.doi.org/10.1140/epjc/s10052-021-09094-9}{\emph{Eur. Phys. J.
  C} {\bf 81} (2021) 320}, [\href{http://arxiv.org/abs/2005.02119}{{\tt
  2005.02119}}].

\bibitem{Plenter:2022zxk}
J.~Plenter, \emph{{Asymptotic expansions and causal representations through the
  loop-tree duality}}.
\newblock PhD thesis, Valencia U., IFIC, 2022.

\bibitem{Driencourt-Mangin:2019aix}
F.~Driencourt-Mangin, G.~Rodrigo, G.~F.~R. Sborlini and W.~J. Torres~Bobadilla,
  \emph{{Universal four-dimensional representation of $H \to \gamma \gamma$ at
  two loops through the Loop-Tree Duality}},
  \href{http://dx.doi.org/10.1007/JHEP02(2019)143}{\emph{JHEP} {\bf 02} (2019)
  143}, [\href{http://arxiv.org/abs/1901.09853}{{\tt 1901.09853}}].

\bibitem{Driencourt-Mangin:2019yhu}
F.~Driencourt-Mangin, G.~Rodrigo, G.~F.~R. Sborlini and W.~J. Torres~Bobadilla,
  \emph{{Interplay between the loop-tree duality and helicity amplitudes}},
  \href{http://dx.doi.org/10.1103/PhysRevD.105.016012}{\emph{Phys. Rev. D} {\bf
  105} (2022) 016012}, [\href{http://arxiv.org/abs/1911.11125}{{\tt
  1911.11125}}].

\bibitem{Hernandez-Pinto:2015ysa}
R.~J. Hernandez-Pinto, G.~F.~R. Sborlini and G.~Rodrigo, \emph{{Towards gauge
  theories in four dimensions}},
  \href{http://dx.doi.org/10.1007/JHEP02(2016)044}{\emph{JHEP} {\bf 02} (2016)
  044}, [\href{http://arxiv.org/abs/1506.04617}{{\tt 1506.04617}}].

\bibitem{Sborlini:2016gbr}
G.~F.~R. Sborlini, F.~Driencourt-Mangin, R.~Hernandez-Pinto and G.~Rodrigo,
  \emph{{Four-dimensional unsubtraction from the loop-tree duality}},
  \href{http://dx.doi.org/10.1007/JHEP08(2016)160}{\emph{JHEP} {\bf 08} (2016)
  160}, [\href{http://arxiv.org/abs/1604.06699}{{\tt 1604.06699}}].

\bibitem{Sborlini:2016hat}
G.~F.~R. Sborlini, F.~Driencourt-Mangin and G.~Rodrigo, \emph{{Four-dimensional
  unsubtraction with massive particles}},
  \href{http://dx.doi.org/10.1007/JHEP10(2016)162}{\emph{JHEP} {\bf 10} (2016)
  162}, [\href{http://arxiv.org/abs/1608.01584}{{\tt 1608.01584}}].

\bibitem{Prisco:2020kyb}
R.~M. Prisco and F.~Tramontano, \emph{{Dual subtractions}},
  \href{http://dx.doi.org/10.1007/JHEP06(2021)089}{\emph{JHEP} {\bf 06} (2021)
  089}, [\href{http://arxiv.org/abs/2012.05012}{{\tt 2012.05012}}].

\bibitem{Aguilera-Verdugo:2019kbz}
J.~J. Aguilera-Verdugo, F.~Driencourt-Mangin, J.~Plenter,
  S.~Ram\'\i{}rez-Uribe, G.~Rodrigo, G.~F.~R. Sborlini et~al.,
  \emph{{Causality, unitarity thresholds, anomalous thresholds and infrared
  singularities from the loop-tree duality at higher orders}},
  \href{http://dx.doi.org/10.1007/JHEP12(2019)163}{\emph{JHEP} {\bf 12} (2019)
  163}, [\href{http://arxiv.org/abs/1904.08389}{{\tt 1904.08389}}].

\bibitem{Verdugo:2020kzh}
J.~J. Aguilera-Verdugo, F.~Driencourt-Mangin, R.~J. Hernandez~Pinto,
  J.~Plenter, S.~Ramirez-Uribe, A.~E. Renteria~Olivo et~al., \emph{{Open loop
  amplitudes and causality to all orders and powers from the loop-tree
  duality}},
  \href{http://dx.doi.org/10.1103/PhysRevLett.124.211602}{\emph{Phys. Rev.
  Lett.} {\bf 124} (2020) 211602}, [\href{http://arxiv.org/abs/2001.03564}{{\tt
  2001.03564}}].

\bibitem{Aguilera-Verdugo:2020kzc}
J.~J. Aguilera-Verdugo, R.~J. Hernandez-Pinto, G.~Rodrigo, G.~F.~R. Sborlini
  and W.~J. Torres~Bobadilla, \emph{{Causal representation of multi-loop
  Feynman integrands within the loop-tree duality}},
  \href{http://dx.doi.org/10.1007/JHEP01(2021)069}{\emph{JHEP} {\bf 01} (2021)
  069}, [\href{http://arxiv.org/abs/2006.11217}{{\tt 2006.11217}}].

\bibitem{Aguilera-Verdugo:2020nrp}
J.~Aguilera-Verdugo, R.~J. Hern\'andez-Pinto, G.~Rodrigo, G.~F.~R. Sborlini and
  W.~J. Torres~Bobadilla, \emph{{Mathematical properties of nested residues and
  their application to multi-loop scattering amplitudes}},
  \href{http://dx.doi.org/10.1007/JHEP02(2021)112}{\emph{JHEP} {\bf 02} (2021)
  112}, [\href{http://arxiv.org/abs/2010.12971}{{\tt 2010.12971}}].

\bibitem{Ramirez-Uribe:2020hes}
S.~Ram\'\i{}rez-Uribe, R.~J. Hern\'andez-Pinto, G.~Rodrigo, G.~F.~R. Sborlini
  and W.~J. Torres~Bobadilla, \emph{{Universal opening of four-loop scattering
  amplitudes to trees}},
  \href{http://dx.doi.org/10.1007/JHEP04(2021)129}{\emph{JHEP} {\bf 04} (2021)
  129}, [\href{http://arxiv.org/abs/2006.13818}{{\tt 2006.13818}}].

\bibitem{Ramirez-Uribe:2022sja}
S.~Ram\'\i{}rez-Uribe, R.~J. Hern\'andez-Pinto, G.~Rodrigo and G.~F.~R.
  Sborlini, \emph{{From Five-Loop Scattering Amplitudes to Open Trees with the
  Loop-Tree Duality}},
  \href{http://dx.doi.org/10.3390/sym14122571}{\emph{Symmetry} {\bf 14} (2022)
  2571}, [\href{http://arxiv.org/abs/2211.03163}{{\tt 2211.03163}}].

\bibitem{Runkel:2019yrs}
R.~Runkel, Z.~Sz\H{o}r, J.~P. Vesga and S.~Weinzierl, \emph{{Causality and
  loop-tree duality at higher loops}},
  \href{http://dx.doi.org/10.1103/PhysRevLett.122.111603}{\emph{Phys. Rev.
  Lett.} {\bf 122} (2019) 111603}, [\href{http://arxiv.org/abs/1902.02135}{{\tt
  1902.02135}}].

\bibitem{Runkel:2019zbm}
R.~Runkel, Z.~Sz\H{o}r, J.~P. Vesga and S.~Weinzierl, \emph{{Integrands of loop
  amplitudes within loop-tree duality}},
  \href{http://dx.doi.org/10.1103/PhysRevD.101.116014}{\emph{Phys. Rev. D} {\bf
  101} (2020) 116014}, [\href{http://arxiv.org/abs/1906.02218}{{\tt
  1906.02218}}].

\bibitem{Capatti:2019edf}
Z.~Capatti, V.~Hirschi, D.~Kermanschah, A.~Pelloni and B.~Ruijl,
  \emph{{Numerical Loop-Tree Duality: contour deformation and subtraction}},
  \href{http://dx.doi.org/10.1007/JHEP04(2020)096}{\emph{JHEP} {\bf 04} (2020)
  096}, [\href{http://arxiv.org/abs/1912.09291}{{\tt 1912.09291}}].

\bibitem{Capatti:2019ypt}
Z.~Capatti, V.~Hirschi, D.~Kermanschah and B.~Ruijl, \emph{{Loop-Tree Duality
  for Multiloop Numerical Integration}},
  \href{http://dx.doi.org/10.1103/PhysRevLett.123.151602}{\emph{Phys. Rev.
  Lett.} {\bf 123} (2019) 151602}, [\href{http://arxiv.org/abs/1906.06138}{{\tt
  1906.06138}}].

\bibitem{TorresBobadilla:2021ivx}
W.~J. Torres~Bobadilla, \emph{{Loop-tree duality from vertices and edges}},
  \href{http://dx.doi.org/10.1007/JHEP04(2021)183}{\emph{JHEP} {\bf 04} (2021)
  183}, [\href{http://arxiv.org/abs/2102.05048}{{\tt 2102.05048}}].

\bibitem{TorresBobadilla:2021dkq}
W.~J.~T. Bobadilla, \emph{{Lotty \textendash{} The loop-tree duality
  automation}},
  \href{http://dx.doi.org/10.1140/epjc/s10052-021-09235-0}{\emph{Eur. Phys. J.
  C} {\bf 81} (2021) 514}, [\href{http://arxiv.org/abs/2103.09237}{{\tt
  2103.09237}}].

\bibitem{Sborlini:2021owe}
G.~F.~R. Sborlini, \emph{{Geometrical approach to causality in multiloop
  amplitudes}},
  \href{http://dx.doi.org/10.1103/PhysRevD.104.036014}{\emph{Phys. Rev. D} {\bf
  104} (2021) 036014}, [\href{http://arxiv.org/abs/2102.05062}{{\tt
  2102.05062}}].

\bibitem{Capatti:2022mly}
Z.~Capatti, \emph{{Exposing the threshold structure of loop integrals}},
  \href{http://dx.doi.org/10.1103/PhysRevD.107.L051902}{\emph{Phys. Rev. D}
  {\bf 107} (2023) L051902}, [\href{http://arxiv.org/abs/2211.09653}{{\tt
  2211.09653}}].

\bibitem{Ramirez-Uribe:2021ubp}
S.~Ram\'\i{}rez-Uribe, A.~E. Renter\'\i{}a-Olivo, G.~Rodrigo, G.~F.~R. Sborlini
  and L.~Vale~Silva, \emph{{Quantum algorithm for Feynman loop integrals}},
  \href{http://dx.doi.org/10.1007/JHEP05(2022)100}{\emph{JHEP} {\bf 05} (2022)
  100}, [\href{http://arxiv.org/abs/2105.08703}{{\tt 2105.08703}}].

\bibitem{Clemente:2022nll}
G.~Clemente, A.~Crippa, K.~Jansen, S.~Ram\'\i{}rez-Uribe, A.~E.
  Renter\'\i{}a-Olivo, G.~Rodrigo et~al., \emph{{Variational quantum
  eigensolver for causal loop Feynman diagrams and acyclic directed graphs}},
  \href{http://arxiv.org/abs/2210.13240}{{\tt 2210.13240}}.

\bibitem{Cutkosky:1960sp}
R.~E. Cutkosky, \emph{{Singularities and discontinuities of Feynman
  amplitudes}}, \href{http://dx.doi.org/10.1063/1.1703676}{\emph{J. Math.
  Phys.} {\bf 1} (1960) 429--433}.

\bibitem{Rodrigo:2023PREP}
G.~Rodrigo et~al., \emph{{in preparation}}.

\bibitem{Sborlini:2021nqu}
G.~F.~R. Sborlini, \emph{{Geometry and causality for efficient multiloop
  representations}},  in \emph{{15th International Symposium on Radiative
  Corrections: Applications of Quantum Field Theory to Phenomenology and
  LoopFest XIX: Workshop on Radiative Corrections for the LHC and Future
  Colliders}}, September 2021.
\newblock \href{http://arxiv.org/abs/2109.07808}{{\tt 2109.07808}}.

\bibitem{Wei:2019rqy}
A.~Y. Wei, P.~Naik, A.~W. Harrow and J.~Thaler, \emph{{Quantum Algorithms for
  Jet Clustering}},
  \href{http://dx.doi.org/10.1103/PhysRevD.101.094015}{\emph{Phys. Rev. D} {\bf
  101} (2020) 094015}, [\href{http://arxiv.org/abs/1908.08949}{{\tt
  1908.08949}}].

\bibitem{Pires:2021fka}
D.~Pires, P.~Bargassa, J.~Seixas and Y.~Omar, \emph{{A Digital Quantum
  Algorithm for Jet Clustering in High-Energy Physics}},
  \href{http://arxiv.org/abs/2101.05618}{{\tt 2101.05618}}.

\bibitem{Pires:2020urc}
D.~Pires, Y.~Omar and J.~Seixas, \emph{{Adiabatic Quantum Algorithm for
  Multijet Clustering in High Energy Physics}},
  \href{http://arxiv.org/abs/2012.14514}{{\tt 2012.14514}}.

\bibitem{deLejarza:2022bwc}
J.~J.~M. de~Lejarza, L.~Cieri and G.~Rodrigo, \emph{{Quantum clustering and jet
  reconstruction at the LHC}},
  \href{http://dx.doi.org/10.1103/PhysRevD.106.036021}{\emph{Phys. Rev. D} {\bf
  106} (2022) 036021}, [\href{http://arxiv.org/abs/2204.06496}{{\tt
  2204.06496}}].

\bibitem{Delgado:2022snu}
A.~Delgado and J.~Thaler, \emph{{Quantum Annealing for Jet Clustering with
  Thrust}},  \href{http://arxiv.org/abs/2205.02814}{{\tt 2205.02814}}.

\bibitem{Barata:2022wim}
J.~a. Barata, X.~Du, M.~Li, W.~Qian and C.~A. Salgado, \emph{{Medium induced
  jet broadening in a quantum computer}},
  \href{http://arxiv.org/abs/2208.06750}{{\tt 2208.06750}}.

\bibitem{Barata:2021yri}
J.~a. Barata and C.~A. Salgado, \emph{{A quantum strategy to compute the jet
  quenching parameter $\hat{q}$}},
  \href{http://dx.doi.org/10.1140/epjc/s10052-021-09674-9}{\emph{Eur. Phys. J.
  C} {\bf 81} (2021) 862}, [\href{http://arxiv.org/abs/2104.04661}{{\tt
  2104.04661}}].

\bibitem{Perez-Salinas:2020nem}
A.~P\'erez-Salinas, J.~Cruz-Martinez, A.~A. Alhajri and S.~Carrazza,
  \emph{{Determining the proton content with a quantum computer}},
  \href{http://dx.doi.org/10.1103/PhysRevD.103.034027}{\emph{Phys. Rev. D} {\bf
  103} (2021) 034027}, [\href{http://arxiv.org/abs/2011.13934}{{\tt
  2011.13934}}].

\bibitem{Ngairangbam:2021yma}
V.~S. Ngairangbam, M.~Spannowsky and M.~Takeuchi, \emph{{Anomaly detection in
  high-energy physics using a quantum autoencoder}},
  \href{http://dx.doi.org/10.1103/PhysRevD.105.095004}{\emph{Phys. Rev. D} {\bf
  105} (2022) 095004}, [\href{http://arxiv.org/abs/2112.04958}{{\tt
  2112.04958}}].

\bibitem{Agliardi:2022ghn}
G.~Agliardi, M.~Grossi, M.~Pellen and E.~Prati, \emph{{Quantum integration of
  elementary particle processes}},
  \href{http://dx.doi.org/10.1016/j.physletb.2022.137228}{\emph{Phys. Lett. B}
  {\bf 832} (2022) 137228}, [\href{http://arxiv.org/abs/2201.01547}{{\tt
  2201.01547}}].

\bibitem{deLejarza:2023qxk}
J.~J.~M. de~Lejarza, M.~Grossi, L.~Cieri and G.~Rodrigo, \emph{{Quantum Fourier
  Iterative Amplitude Estimation}},
  \href{http://arxiv.org/abs/2305.01686}{{\tt 2305.01686}}.

\bibitem{Grover:1997ch}
L.~K. Grover, \emph{{Quantum computers can search rapidly by using almost any
  transformation}},
  \href{http://dx.doi.org/10.1103/PhysRevLett.80.4329}{\emph{Phys. Rev. Lett.}
  {\bf 80} (1998) 4329--4332},
  [\href{http://arxiv.org/abs/quant-ph/9712011}{{\tt quant-ph/9712011}}].

\bibitem{Grover:1997fa}
L.~K. Grover, \emph{{Quantum mechanics helps in searching for a needle in a
  haystack}}, \href{http://dx.doi.org/10.1103/PhysRevLett.79.325}{\emph{Phys.
  Rev. Lett.} {\bf 79} (1997) 325--328},
  [\href{http://arxiv.org/abs/quant-ph/9706033}{{\tt quant-ph/9706033}}].

\bibitem{Tilly:2021jem}
J.~Tilly et~al., \emph{{The Variational Quantum Eigensolver: A review of
  methods and best practices}},
  \href{http://dx.doi.org/10.1016/j.physrep.2022.08.003}{\emph{Phys. Rept.}
  {\bf 986} (2022) 1--128}, [\href{http://arxiv.org/abs/2111.05176}{{\tt
  2111.05176}}].

\end{thebibliography}

\providecommand{\href}[2]{#2}\begingroup\raggedright\endgroup

\end{document}